\documentstyle[aps,prb,epsf,multicol]{revtex}
\begin{document}

\newcommand{\news}{{\tilde s}}
\newcommand{\newz}{{\tilde z}}

\title{Doped bilayer antiferromagnets: Hole dynamics on both sides of a
       magnetic ordering transition}
\author{Matthias Vojta$^{(a)}$ and Klaus W. Becker$^{(b)}$}
\address{
  (a) Department of Physics and Applied Physics,
      Yale University,
      New Haven, CT 06520, USA \\
  (b) Institut f\"{u}r Theoretische Physik,
      Technische Universit\"{a}t Dresden, D-01062 Dresden, Germany
}
\maketitle

\begin{abstract}
The two-layer square lattice quantum antiferromagnet with spins $1 \over 2$ shows a
magnetic order-disorder transition at a critical ratio of the interplane to
intraplane couplings.
We investigate the dynamics of a single hole in a bilayer antiferromagnet described by
a $t-J$ Hamiltonian.
To model the spin background we propose a ground-state wave function for the undoped system
which covers both magnetic phases and includes transverse as well as longitudinal
spin fluctuations.
The photoemission spectrum is calculated using the spin-polaron picture
for the whole range of the ratio of the magnetic couplings.
This allows for the study of the hole dynamics of both sides of the magnetic
order-disorder transition.
For small interplane coupling we find a quasiparticle with properties
known from the single-layer antiferromagnet, e.g., the dispersion minimum
is at $(\pm\pi/2,\pm\pi/2)$.
For large interplane coupling the hole dispersion is similar to that of a free
fermion (with reduced bandwidth).
The cross-over between these two scenarios occurs inside the antiferromagnetic
phase which indicates that the hole dynamics is governed by the local
environment of the hole.
\end{abstract}

\widetext
\begin{multicols}{2}
\narrowtext


\section{Introduction}

Since the discovery of high-temperature superconductivity,
doped antiferromagnets (AF) have been studied intensively.
It is widely accepted that many properties of the superconducting
cuprates are determined by the hole-doped CuO$_2$ planes.
A number of experiments \cite{AFScale} indicate that the cuprates are near a quantum-critical
point of antiferromagnetic instability:
The undoped materials are known to be antiferromagnetic Mott-Hubbard
insulators, whereas hole doping destroys the antiferromagnetic
long-range order (LRO) and leads to superconductivity.
The investigation of the interplay between long-range magnetic order
and quantum disorder is therefore of great theoretical interest.

A model system which shows a quantum transition between an ordered
and a disordered magnetic phase is the $S={1 \over 2}$ bilayer antiferromagnet
\cite{Chub95,SaSca,SaChuSa,Duin97,Zheng97,Kotov98,Hida90,MillMo}.
Here each of the two planes is composed of a nearest-neighbor Heisenberg model
with coupling constant $J_\parallel$. The spins of corresponding
sites of each layer are coupled antiferromagnetically with a coupling constant
$J_\perp$.
In the limit of small $J_\perp/J_\parallel$ the model describes two weakly
coupled AF planes. At $T=0$ this system possesses AF long-range order and gapless
Goldstone excitations.
In the opposite case of large $J_\perp/J_\parallel$, pairs of spins interacting
via $J_\perp$ form spin singlets being weakly coupled by $J_\parallel$.
Then the spin excitations are gapped triplet modes, there is no magnetic
LRO.
At a critical ratio $(J_\perp/J_\parallel)_c$ a quantum transition between
the two phases occurs which is believed to be of the O(3) universality
class \cite{Chub95,SaSca,SaChuSa,Duin97}.
The bilayer antiferromagnet has been studied by a number of numerical
and analytical techniques.
Quantum Monte Carlo calculations \cite{SaSca,SaChuSa} and series expansions
\cite{Zheng97} yield an order-disorder transition point of
$(J_\perp/J_\parallel)_c \sim 2.5$.
A similar result has also been obtained analytically using a diagrammatic
approach to account for the hard-core interaction between triplet
excitations \cite{Kotov98}.
Bond-operator mean-field theory has recently been applied to the
bilayer Heisenberg AF \cite{BO-MF} and gives a transition point of
$(J_\perp/J_\parallel)_c \sim 2.3$.
Note that Schwinger boson mean-field theory \cite{MillMo} predicts a
very large value of $(J_\perp/J_\parallel)_c \sim 4.5$, and also
self-consistent spin-wave theory \cite{Chub95,Hida90},
which yields $(J_\perp/J_\parallel)_c \sim 4.3$, fails to
reproduce the numerical results. Chubukov and Morr \cite{Chub95}
have pointed out
that this discrepancy is due to the neglect of longitudinal
spin fluctuations in the conventional spin-wave approach.

In this paper we discuss the bilayer antiferromagnet at zero temperature
with hole doping as an additional degree of freedom.
We consider the standard $t-J$ model on a bilayer square lattice consisting
of $N$ sites per plane, so the total number of lattice sites
is $2N$.
Each pair of corresponding sites in different planes is considered to form a rung,
so we have $N$ rungs. More precisely, the Hamiltonian reads
\begin{eqnarray}
H =
&-& t_\perp\; \sum_{i\sigma}
(\hat{c}_{i1,\sigma}^\dagger \hat{c}_{i2,\sigma}^{} + H.c.)
\nonumber \\
&-& t_\parallel\; \sum_{\langle ij\rangle m\sigma}
(\hat{c}_{im,\sigma}^\dagger \hat{c}_{jm,\sigma}^{} + H.c.)
\nonumber \\
&+& J_\perp \; \sum_{i}
(\bbox{S}_{i1} \cdot \bbox{S}_{i2} \,-\, {n_{i1} n_{i2} \over 4})
\nonumber \\
&+& J_\parallel \; \sum_{\langle ij\rangle m}
(\bbox{S}_{im} \cdot \bbox{S}_{jm} \,-\, {n_{im} n_{jm} \over 4})
\;.
\label{H_TJ}
\end{eqnarray}
Here and in the following $i$ and $j$ denote $rungs$ of the bilayer lattice,
$\langle ij\rangle$ denotes a summation over all pairs of
nearest-neighbor rungs.
$m=1,2$ is the plane index, so $im$ denotes a lattice site.
${\bf S}_{im}$ is the electronic spin operator and $n_{im}$ the electron
number operator at site $im$.
The electron operators $\hat c_{im,\sigma}^{\dagger}$ exclude double occupancies,
\mbox{$\hat{c}_{im,\sigma}^\dagger = {c}_{im,\sigma}^\dagger (1-n_{im,-\sigma})$}.
We choose the $z$-axis along the direction of the rungs, i.e., perpendicular
to the planes.

First we want to comment the zero-temperature phase diagram
of the doped bilayer $t-J$ model which has to our knowledge
not been systematically studied up to now.
The following non-thermal control parameters can be considered:
the ratio $J_\perp/J_\parallel$, the doping level $\delta$,
and the relative hopping strength $t/J$.
The doped single-layer antiferromagnet is known to exhibit a
strong dependence of magnetic properties on the hole concentration $\delta$.
With increasing hole concentration the staggered magnetization
decreases and vanishes at a critical hole
concentration $\delta_c$ of a few percent
where the system becomes paramagnetic
\cite{Dagotto94,Brenig95,Fulde,VojBeck96}.
(This is consistent with experiments on high-$T_c$
superconductors \cite{Dagotto94,Brenig95}.)
On the other hand, in the undoped limit a large interplane coupling $J_\perp$
also destroys the antiferromagnetic LRO as discussed above.
Thus it is likely that the system shows (i) AF LRO accompanied by spontaneous symmetry
breaking and the existence of massless Goldstone modes
in the region of very small doping and small $J_\perp$.
Otherwise it is paramagnetic;
here one has to distinguish between (ii) a spin-gapped phase which occurs
at small $\delta$, small hopping $t_\parallel$, and large interplane
coupling,
and (iii) a gapless phase at larger doping and small interplane coupling.
The gapped phase (ii) is dominated by interplane singlet pairs
(incompressible spin fluid) whereas the gapless phase (iii) shows
only weak interplane coupling and behaves mainly like a single layer
system at $\delta>\delta_c$ (compressible spin fluid).
(Issues like pseudogap features and transport properties will not be adressed
here.)
In the present work we restrict ourselves to the zero-doping limit
(one hole).
So the magnetic (bulk) properties will not be affected by the doped hole,
i.e., we probe the transition between the antiferromagnetic (i)
and the gapped (incompressible) paramagnetic phase (ii).

The dynamical properties of holes in high-$T_c$ superconductors have
been subject to a lot of experimental and theoretical work.
Angle-resolved photoemission (ARPES) experiments
at zero\cite{Wells95} or small doping\cite{Dessau93,Marshall96}
indicate a quasiparticle band with a small bandwidth providing
evidence for strong electronic correlations in the
high-$T_c$ compounds.
A large number of numerical and analytical studies reveal that
a single hole in an AF spin background has non-trivial
properties:
The spectral function consists of a pronounced coherent peak at the bottom of the
spectrum and a incoherent background at higher energies.
The coherent peak can be associated with the motion of a dressed hole,
i.e., a hole surrounded by spin defects ("spin polaron")
~\cite{Dagotto94,Brenig95,Strings,Trugman88,Schrieffer,MaHo,Riera97,Vojta98}.
The hole motion in bilayer antiferromagnets has been studied in a few
papers and only for a small parameter range:
Using self-consistent Born approximation (SCBA) it has been shown \cite{BilaSCBA} that
small interplane coupling only weakly affects the hole properties.
Exact diagonalization studies \cite{EderBila} have indicated a larger
effect of the interplane coupling at finite doping.
However, the system sizes accessible to numerical methods
($2\times 8$ in Ref. \onlinecite{EderBila}) do not allow for the study
of small hole densities and arbitrary momenta.

In this paper we prefer an analytical approach to the one-hole problem
which is based on the
picture of the spin-bag quasiparticle (QP) or magnetic polaron
\cite{Strings,Trugman88,Schrieffer,MaHo,Riera97,Vojta98}.
The magnetic background for the hole motion will be modelled
within a modified bond-operator representation of the spins
on each rung.
In the ordered phase one type of triplet bosons condenses which
will be included in the modified basis operators.
So our excitation operators continuously interpolate between
the triplet excitations of a singlet product state and
the transverse and longitudinal excitations of a N\'{e}el-ordered
state.
The deviations of the spin background caused by the hole motion
are described by a set of path operators
\cite{Strings,Trugman88,Riera97,Vojta98}.
The one-hole spectral function will be evaluated using
a cumulant version \cite{BeckFul88,BeckBre90}
of Mori-Zwanzig projection technique\cite{Projtech}.

The paper is organized as follows:
In Sec. II we propose a ground-state wave function for the undoped
bilayer antiferromagnet.
In Sec. III we develop the Hamiltonian for the doped system in
this new operator representation and discuss the various hole motion
processes.
Sec. IV focuses on the motion of
a single hole in an otherwise half-filled bilayer antiferromagnet.
The one-hole spectral function for this case corresponds directly to the
result of an ARPES experiment on an undoped system.
After a brief sketch of the projection technique
we investigate the spectral properties in dependence of the
magnetic background, especially
when crossing the phase boundary between gapped paramagnet and
antiferromagnet.
The results show that the character of the one-hole spectrum is mainly
determined by the local spin environment of the hole
whereas the spectrum changes only weakly when crossing the order-disorder
transition.
A conclusion will close the paper.


\section{Undoped bilayer antiferromagnet}

To begin with, we consider the case of half-filling.
Using bond operators we derive a wave function which describes
the ground states in
both the quantum disordered phase (at large $J_\perp$) and
the N\'{e}el phase (at small $J_\perp$).
The wave function will be constructed within a cumulant formalism
\cite{BeckFul88,SchorkFul92}.
It bases on an expansion around a product of rung states
which are singlets in the disordered phase and sums of singlets
and $z$-type triplets in the ordered phase.
This wave function will be used in Sec. IV as "magnetic background" to
calculate dynamical properties of a single hole.

\subsection{Rung basis and Hamiltonian}

In the limit of vanishing intraplane coupling, $J_\parallel/J_\perp \rightarrow 0$,
pairs of spins on each rung form a singlet ground state. The excitations are
localized triplets with an energy gap $J_\perp$.
Switching on $J_\parallel$ leads to an interaction of triplets on neighboring rungs
and to a dispersion of the triplet excitations.
To describe this dimerized phase we employ a bond operator representation \cite{SaBha} for
the spins per rung.
For each rung $i$ containing two $S=1/2$ spins ${\bf S}_{i1}$,
${\bf S}_{i2}$ we introduce operators for creation
of a singlet and three triplet states out of the vacuum $|0\rangle$:
\end{multicols}
\widetext
\begin{eqnarray}
s_i^{\dagger}|0\rangle \:&=&\:
  \frac{1}{\sqrt{2}}
    (\hat{c}_{i1,\uparrow}^\dagger \hat{c}_{i2,\downarrow}^\dagger
    - \hat{c}_{i1,\downarrow}^\dagger \hat{c}_{i2,\uparrow}^\dagger) |0\rangle \:=\:
  \frac{1}{\sqrt{2}}(|\uparrow\downarrow\rangle - |\downarrow \uparrow\rangle)
\:,
\nonumber\\
t_{ix}^{\dagger}|0\rangle \:&=&\:
  \frac{-1}{\sqrt{2}}(
    \hat{c}_{i1,\uparrow}^\dagger \hat{c}_{i2,\uparrow}^\dagger
    -\hat{c}_{i1,\downarrow}^\dagger \hat{c}_{i2,\downarrow}^\dagger) |0\rangle \:=\:
  \frac{-1}{\sqrt{2}} (|\uparrow \uparrow\rangle - |\downarrow \downarrow\rangle)
\:,
\nonumber\\
t_{iy}^{\dagger}|0\rangle \:&=&\:
  \frac{i}{\sqrt{2}}(
    \hat{c}_{i1,\uparrow}^\dagger \hat{c}_{i2,\uparrow}^\dagger +
    \hat{c}_{i1,\downarrow}^\dagger \hat{c}_{i2,\downarrow}^\dagger)  |0\rangle \:=\:
  \frac{i}{\sqrt{2}}(|\uparrow \uparrow\rangle + |\downarrow \downarrow\rangle)
\:,
\nonumber\\
t_{iz}^{\dagger}|0\rangle \:&=&\:
  \frac{1}{\sqrt{2}}
    (\hat{c}_{i1,\uparrow}^\dagger \hat{c}_{i2,\downarrow}^\dagger
    + \hat{c}_{i1,\downarrow}^\dagger \hat{c}_{i2,\uparrow}^\dagger)  |0\rangle \:=\:
  \frac{1}{\sqrt{2}}(|\uparrow \downarrow\rangle + |\downarrow \uparrow\rangle).
\label{TRIP_DEF}
\end{eqnarray}
\begin{multicols}{2}
\narrowtext
Then the following representation holds \cite{SaBha}:
\begin{equation}
S_{i1,2}^{\alpha} = \frac{1}{2} ( \pm s_i^{\dagger}  t_{i\alpha}^{} \pm
t_{i\alpha}^{\dagger} s_i^{}  - i \epsilon_{\alpha\beta\gamma} t_{i\beta}^{\dagger}
t_{i\gamma})
\,.
\label{SPIN_REP}
\end{equation}
The algebra of the operators $\{s_i,t_{ix},t_{iy},t_{iz}\}$ has to be specified to
reproduce the correct algebra for the spin operators.
Here bond operators satisfying the usual bosonic commutation relations will
be chosen \cite{SaBha}.
In order to ensure that the physical states are either singlets or
triplets one has to impose the condition:
\begin{equation}
s_i^{\dagger} s_i^{} + \sum_\alpha t_{i\alpha}^{\dagger}t_{i\alpha}^{} = 1.
\label{CONSTRAINT}
\end{equation}
The singlet product ground state for $J_\parallel/J_\perp \rightarrow 0$
can be written as
\begin{eqnarray}
|\phi_0\rangle \:=\: \prod_i s_i^\dagger |0\rangle \:.
\label{SINGLET_PROD}
\end{eqnarray}

We can now substitute the operator representation (\ref{SPIN_REP})
into the Heisenberg Hamiltonian for the bilayer
and obtain the following Hamiltonian $H=H_0+H_1$ in the bond (rung) operator
representation \cite{Gopalan,Eder98}:
\begin{eqnarray}
H_0 \:&=&\: J_\perp \sum_{i\alpha}
t_{i\alpha}^\dagger t_{i\alpha}^{} \:,
\nonumber \\
H_1 \:&=&\: \frac{J_\parallel}{2} \sum_{\langle ij\rangle\alpha}
(\;t_{i\alpha}^\dagger t_{j\alpha}^\dagger \; s_j^{} s_i^{} + H.c.\;)
\nonumber \\
&+&\frac{J_\parallel}{2} \sum_{\langle ij\rangle\alpha}
(\; t_{i\alpha}^\dagger s_j^\dagger \; t_{j\alpha}^{} s_i^{} + H.c.\;)
\nonumber \\
&-&\frac{J_\parallel}{2}
\sum_{\langle ij\rangle\alpha\beta} (\;
  t_{i\alpha}^\dagger t_{j\alpha}^\dagger\;
  t_{j\beta}^{} t_{i\beta}^{}
- t_{i\alpha}^\dagger t_{j\beta}^\dagger \;
  t_{j\alpha}^{} t_{i\beta}^{}\;).
\label{H_UNDOP_DISO}
\end{eqnarray}
The sum $\langle ij\rangle$ runs over pairs of neighboring rungs
of the lattice.
The ground state of $H_0$ is given by the product state $|\phi_0\rangle$
defined in eq. (\ref{SINGLET_PROD}).

Approaching the quantum critical point the triplet excitations will become gapless
at momentum $(\pi,\pi)$, see e.g. Refs. \onlinecite{Chub95,Zheng97}.
The magnetic quantum phase transition
to an antiferromagnetically ordered state includes spontaneous symmetry
breaking and can be described via the condensation of triplet bosons
in one particular direction (which has to be induced by an infinitesimal
staggered field).
Neglecting fluctuations a symmetry-broken state with a condensate of $z$-type triplet bosons
can be written as
\begin{eqnarray}
&&|\tilde \phi_0\rangle \: \sim \:
\nonumber \\
&&\exp(\lambda \sum_i {\rm e}^{{\rm i} {\bf Q R}_i} t_{iz}^\dagger) |\phi_0 \rangle
\:=\: \prod_i (s_i^\dagger + \lambda {\rm e}^{{\rm i} {\bf Q R}_i} t_{iz}^\dagger) |0\rangle
\label{PHI_SYMMB}
\end{eqnarray}
where the spontaneous staggered magnetization points in $z$-direction.
The exponential term accounts for boson condensation, the
condensation amplitude is given
by $\lambda$.
$\lambda=0$ means rotational symmetry in spin space (then $|\tilde \phi_0\rangle =
|\phi_0\rangle$),
$\lambda\neq 0$ breaks this symmetry.
$\lambda=\pm 1$ transforms the singlet product state $|\phi_0\rangle$
into a N\'{e}el state.
$\bf Q = (\pi,\pi)$ is the AF ordering wave vector.

The basic idea for a proper description of the excitations of the product state
$|\tilde \phi_0\rangle$ (\ref{PHI_SYMMB})
is now to transform the basis states
on each rung. We replace the basis operators $s_i$ for singlets and $t_{iz}$ for
$z$-triplets by
\begin{eqnarray}
\news_i   \:&=&\: {1 \over \sqrt{1+\lambda^2} } \:
  (s_i + \lambda {\rm e}^{{\rm i} {\bf Q R}_i} t_{iz}) \,,
\nonumber \\
{\newz}_i \:&=&\: {1 \over \sqrt{1+\lambda^2} } \:
  (- \lambda {\rm e}^{{\rm i} {\bf Q R}_i} s_i + t_{iz}) \,.
\label{BASIS_DEF}
\end{eqnarray}
The new basis per rung consists of the operators
$\{\news_i,\newz_i,t_{ix},t_{iy}\}$ which still satisfy bosonic
commutation relations.
For $\lambda=0$ this basis reproduces the usual bond-boson basis.
For non-zero $\lambda$ the state $\news_i^\dagger |0\rangle$ interpolates between the rung singlet and
a (local) N\'{e}el state, the state $\newz_i^\dagger |0\rangle$ is orthogonal
to $\news_i^\dagger |0\rangle$.
The product state $|\tilde \phi_0\rangle$ (\ref{PHI_SYMMB})
can now be written as
\begin{eqnarray}
|\tilde\phi_0\rangle \:=\: \prod_i \news_i^\dagger |0\rangle \:.
\label{CONDENS_PROD}
\end{eqnarray}
For $|\lambda|=1$ this state is an antiferromagnetic N\'{e}el state.
Its excitations are given by $t_{ix}^\dagger \news_i^{}$, $t_{iy}^\dagger \news_i^{}$,
and $\newz_i^\dagger \news_i^{}$.
Here, $(t_{ix}^\dagger \pm i t_{iy}^\dagger) \news_i^{}$ create one spin flip in one of the planes and
can be regarded as transverse spin fluctuations.
The operator $\newz_i^\dagger \news_i^{}$ flips both spins on one rung which can be interpreted as
longitudinal spin fluctuation.
The basis parameter $\lambda$ has to be determined separately which will
be discussed in the next subsection.
Below it is shown that in the ordered phase $\lambda$ varies continuously between
$\lambda=1$ for $J_\perp=0$ and $\lambda=0$ at the transition to the disordered
phase.
Note that the idea of a condensate for one type of bosons has also been
employed by Chubukov and Morr \cite{Chub95} who applied a modified version of non-linear
spin-wave theory to the problem of the undoped bilayer antiferromagnet.

Using (\ref{BASIS_DEF}) we can reformulate the Hamiltonian for the (undoped)
bilayer Heisenberg model in terms of the new basis operators $\{\news_i,\newz_i,t_{ix},t_{iy}\}$.
Note that the splitting of the Hamiltonian is different from (\ref{H_UNDOP_DISO}).
\end{multicols}
\widetext
\begin{eqnarray}
\tilde H_0 &=& J_\perp \sum_i
  \left ( t_{ix}^\dagger t_{ix}^{} + t_{iy}^\dagger t_{iy}^{} +
    {1 \over {1 + \lambda^2 } }
    \left ( \newz_i^\dagger \newz_i^{} +
    \lambda^2 \news_i^\dagger \news_i^{}
    \right )
  \right ) \,,
\nonumber \\
\tilde H_1 &=& J_\perp \sum_i
    {\lambda \over {1 + \lambda^2 } }
    {\rm e}^{{\rm i} {\bf Q R}_i}
    (\newz_i^\dagger \news_i^{} + \news_i^\dagger \newz_i^{})
\nonumber \\
&+&\frac{J_\parallel}{2} \sum_{\langle i\in A, j\in B\rangle}
  \left (
  \;(t_{ix}^\dagger t_{jx}^\dagger + t_{iy}^\dagger t_{jy}^\dagger)
    (\news_i^{} \news_j^{} - \newz_i^{} \newz_j^{} )  + H.c.\;
  \right )
\nonumber \\
&+&\frac{J_\parallel}{2} \sum_{\langle i\in A, j\in B\rangle}
(\;t_{ix}^\dagger t_{jx}^\dagger t_{iy}^{} t_{jy}^{}
 - t_{ix}^\dagger t_{jy}^\dagger t_{iy}^{} t_{jx}^{} + H.c.\;)
\nonumber \\
&+&\frac{J_\parallel}{2 (1+\lambda^2)} \sum_{\langle i\in A, j\in B\rangle}
  \left (
  \;t_{ix}^\dagger \news_j^\dagger
  \left ( (1-\lambda^2) \news_i^{} t_{jx}^{} - 2\lambda \newz_i^{} t_{jx}^{} \right ) +
  \;t_{iy}^\dagger \news_j^\dagger
  \left ( (1-\lambda^2) \news_i^{} t_{jy}^{} - 2\lambda \newz_i^{} t_{jy}^{} \right ) + H.c.\;
  \right)
\nonumber \\
&+&\frac{J_\parallel}{2 (1+\lambda^2)} \sum_{\langle i\in A, j\in B\rangle}
  \left (
  \;t_{ix}^\dagger \newz_j^\dagger
  \left ( (1-\lambda^2) \newz_i^{} t_{jx}^{} + 2\lambda \news_i^{} t_{jx}^{} \right ) +
  \;t_{iy}^\dagger \newz_j^\dagger
  \left ( (1-\lambda^2) \newz_i^{} t_{jy}^{} + 2\lambda \news_i^{} t_{jy}^{} \right ) + H.c.\;
  \right )
\nonumber \\
&+&\frac{J_\parallel}{2 (1+\lambda^2)^2}
\sum_{\langle i\in A, j\in B\rangle}
  4 \lambda^2 (-\news_i^\dagger \news_j^\dagger \news_j^{} \news_i^{}
               -\newz_i^\dagger \newz_j^\dagger \newz_j^{} \newz_i^{}
               +\newz_i^\dagger \news_j^\dagger \news_j^{} \newz_i^{}
               +\news_i^\dagger \newz_j^\dagger \newz_j^{} \news_i^{} )
\nonumber \\
&+&\frac{J_\parallel}{2 (1+\lambda^2)^2}
\sum_{\langle i\in A, j\in B\rangle}
  \left (
  (1-\lambda^2)^2 ( \newz_i^\dagger \newz_j^\dagger \news_i^{} \news_j^{}
                   +\newz_i^\dagger \news_j^\dagger \news_i^{} \newz_j^{} ) +
  2\lambda(1-\lambda^2) (\news_i^\dagger \newz_j^\dagger - \newz_i^\dagger \news_j^\dagger)
                         \news_i^{} \news_j^{}  + H.c.\;
  \right )
\,.
\label{H_UNDOP_O}
\end{eqnarray}
\begin{multicols}{2}
\narrowtext
\noindent
$A$ and $B$ denote the sublattices of the square lattice, i.e,
${\rm e}^{{\rm i} {\bf Q R}_i}= \pm 1$ for $i\in A$ or $B$,
respectively.
For $\lambda=0$ the Hamiltonian (\ref{H_UNDOP_O}) is identical to (\ref{H_UNDOP_DISO}) derived
with the usual bond operators.
For $|\lambda| \leq 1$ the ground state of $\tilde H_0$ defined in (\ref{H_UNDOP_O}) is given by
the modified product state $|\tilde\phi_0\rangle$ (\ref{CONDENS_PROD}).
It can be seen that $\tilde H_1$ in (\ref{H_UNDOP_O}) contains creation,
hopping, and conversion terms of the three types of excitations
$\{\newz_i,t_{ix},t_{iy}\}$.
Contrary to the Hamiltonian $H_1$ in (\ref{H_UNDOP_DISO}) where the triplet excitations
can only occur in pairs, here also single $\newz$-type excitations
can be created and destroyed, e.g., by
$J_\perp (\newz_i^\dagger \news_i^{} + \news_i^\dagger \newz_i^{})$.
The effect of these terms is directly related to the basis parameter
$\lambda$ and will be used to determine its value.
Furthermore it can be seen that all terms creating $\newz$-type excitations
out of $|\tilde \phi_0\rangle$ vanish for $J_\perp = 0, |\lambda| = 1$.
This means that the ground state in the limit of vanishing interplane coupling
$J_\perp$ does not contain single longitudinal spin fluctuations.
They will become, however, important for describing the destruction of the
antiferromagnetism with increasing $J_\perp$.

Note that $\newz$-type bosons formally occur together with transverse fluctuations
even at $J_\perp = 0, |\lambda| = 1$.
This happens if a transverse fluctuation created by $J_\parallel$ hits a second transverse
fluctuation being already present at the same rung, in other words, if spins on corresponding sites
in each of the planes are flipped independently by $J_\parallel$.
These processes have to be included in an exact treatment of the ground-state wave function;
within the approximation of independent bosons (which will be described in Sec. II C)
they are neglected.

\subsection{Approximation for the ground state}

To obtain an approximation for the ground-state wave function we
adopt a cumulant approach \cite{BeckFul88} which has been successfully
applied to a variety of strongly correlated systems.
One starts from a splitting of the Hamiltonian into $H=H_0+H_1$ where
eigenstates and eigenvalues of $H_0$ are known.
The ground state $|\psi_0\rangle$ of $H$ can be constructed from the unperturbed
ground state $|\phi_0\rangle$ of $H_0$ by application of a so-called
wave operator $\Omega$ which contains the effect of $H_1$.
It has been shown \cite{SchorkFul92} that $\Omega$ can be written in
an exponential form, $\Omega=e^S$, where $S$ introduces fluctuations
into $|\phi_0\rangle$.
The ground-state energy $E$ is calculated \cite{BeckFul88} according
to $E = \langle\phi_0 |H \Omega|\phi_0\rangle^c$ where
$\langle\phi_0 | ... |\phi_0\rangle^c$ denotes a cumulant expectation
value with respect to $|\phi_0\rangle$.

For the disordered phase of the system it is suitable to start from the
singlet product state $|\phi_0\rangle$ given in eq. (\ref{SINGLET_PROD})
and to include fluctuations in form of pairs of triplet excitations.
An appropriate ansatz for the wave function is
\begin{eqnarray}
|\psi_0\rangle \:&=&\: \Omega\, |\phi_0 \rangle
\nonumber\\
\:&=&\:
 \exp \left(\sum_{ij\alpha} \mu_{ij\alpha} t_{i\alpha}^\dagger t_{j\alpha}^\dagger s_j^{} s_i^{}
    + \sum_n \beta_n S_n\right)
 |\phi_0\rangle
\,.
\label{OMEGA_DISO}
\end{eqnarray}
The first term in the exponential contains pairs of triplets (with arbitrary
distance \mbox{${\bf R}_i-{\bf R}_j$}).
The operators $S_n$ in the second term represent higher-order
fluctuations, e.g., four-triplet operators.
They will be used in Sec. IV.
For the determination of the coefficients $\mu_{ij\alpha}$ and $\beta_n$ the cumulant
method provides a set of non-linear equations \cite{SchorkFul92},
\begin{eqnarray}
0 &=& \langle\phi_0| (s_i^\dagger s_j^\dagger t_{j\alpha} t_{i\alpha})^\cdot \, H \, \Omega\, |\phi_0\rangle^c
\,,\nonumber\\
0 &=& \langle\phi_0| S_n^\dagger\, H \, \Omega\, |\phi_0\rangle^c \,.
\label{COEFF_EQ_DISO}
\end{eqnarray}
The dot $\cdot$ indicates that the quantity inside $(...)^{\cdot}$
has to be treated as a single entity in the cumulant formation.
The equations (\ref{COEFF_EQ_DISO}) are derived from the requirement that $\Omega |\phi_0\rangle$
is an eigenstate of $H$.
From symmetry it follows that the coefficients $\mu_{ij\alpha} = \mu_{ij}$ only depend
on the difference vector ${\bf R}_i-{\bf R}_j$ (but not on $\alpha$).
So the state $|\psi_0\rangle$ obeys rotational invariance in spin space.
When evaluating the terms in (\ref{COEFF_EQ_DISO}) the exponential series terminates after a few
terms since the exponent contains only operators which create (but not destroy) fluctuations.
The resulting non-linear coupled equations for $\mu_{ij}$ can be partially decoupled by a
Fourier transformation to momentum space.
The remaining set of equations has to be solved self-consistently.

For the ordered phase the treatment is similar, however,
one has to take care of the broken rotational symmetry.
Starting point is the product wave function $|\tilde\phi_0\rangle$ (\ref{CONDENS_PROD}).
In the operator $\Omega$ we now explicitely distinguish between pairs of transverse
($t_{ix}^\dagger t_{jx}^\dagger, t_{iy}^\dagger t_{jy}^\dagger$) and longitudinal
($\newz_{i}^\dagger \newz_{j}^\dagger$) fluctuations:
\end{multicols}
\widetext
\begin{eqnarray}
|\psi_0\rangle \:=\: \Omega\, |\tilde \phi_0 \rangle \:=\:
 \exp \left(
   \sum_{ij} \mu_{ij} (t_{ix}^\dagger t_{jx}^\dagger + t_{iy}^\dagger t_{jy}^\dagger) \news_j^{} \news_i^{} +
   \sum_{ij} \nu_{ij} \newz_{i}^\dagger \newz_{j}^\dagger \news_j^{} \news_i^{}
   + \sum_n \beta_n S_n \right)
 |\tilde \phi_0\rangle
\,.
\label{OMEGA_O}
\end{eqnarray}
\begin{multicols}{2}
\narrowtext
\noindent
The operators $S_n$ again contain higher-order fluctuations and will be discussed later.
The parameters $\mu_{ij}$, $\nu_{ij}$, and $\beta_n$ have to be determined in analogy to
(\ref{COEFF_EQ_DISO}) by the equations
\begin{eqnarray}
0 &=& \langle \tilde \phi_0| (\news_i^\dagger \news_j^\dagger t_{jx} t_{ix})^\cdot \, H \, \Omega\, | \tilde\phi_0\rangle^c
\,,\nonumber\\
0 &=& \langle\ \tilde \phi_0| (\news_i^\dagger \news_j^\dagger \newz_j \newz_i)^\cdot \, H \, \Omega\, |\tilde\phi_0\rangle^c
\,,\nonumber\\
0 &=& \langle \tilde \phi_0| S_n^\dagger\, H \, \Omega\, | \tilde\phi_0\rangle^c
\label{COEFF_EQ_O}
\end{eqnarray}
which again reduce to (\ref{COEFF_EQ_DISO}) in the case of $\lambda=0$.
For $\lambda=0$ we expect the transverse and longitudinal fluctuations to become equivalent,
$\mu_{ij} = \nu_{ij}$.
For $|\lambda|=1$ the amplitude of the longitudinal fluctuations
in the ground state vanishes, $\nu_{ij}=0$, if one neglects higher-order processes in
$J_\parallel$ as noted above.
Note that $\Omega$ in (\ref{OMEGA_O}) does not contain operators for the creation of single
$\newz$-type excitations $(\newz_i^\dagger\news_i^{})$ although such operators are present
in the Hamiltonian (\ref{H_UNDOP_O}).
The reason is that such operators change the density
of the condensate of the $z$ triplet bosons and refer to the same degree of freedom
as the parameter $\lambda$ in the product state $|\tilde\phi_0\rangle$.
The aim here is, however, to describe the condensation of the $z$-type bosons completely via
the introduction of $\lambda$ and the transformed basis states.
In this way $\Omega$ contains only multi-spin fluctuations.

Now we discuss the parameter $\lambda$ which is expected to be zero throughout the disordered
phase and to vary continuously from 0 to 1 with decreasing $J_\perp$ in the ordered phase.
Within the cumulant method $\lambda$ can be determined consistently using
the identity
\begin{equation}
0 = \langle\tilde\phi_0| (\news_i^\dagger\newz_i^{})^\cdot\, H \, \Omega\, |\tilde\phi_0\rangle^c \,.
\label{LAMBDA_EQ}
\end{equation}
This can be understood as the condition that $\Omega$ contains {\bf no}
additional term like $\sum_i \newz_i^\dagger\news_i^{}$ in the exponential,
i.e., the boson condensation is fully accounted for by the transformed basis
state $|\tilde\phi_0\rangle$.
Within the cumulant formalism this can be seen from the equality
\begin{eqnarray}
\Omega\, |\tilde\phi_0\rangle^c \:&=&\:
\Omega\, |e^{\lambda t_{{\bf Q}z}^\dagger} \phi_0\rangle^c \:=\:
\Omega\, e^{\lambda t_{{\bf Q}z}^\dagger} |\phi_0\rangle^c \,, \nonumber\\
t_{{\bf Q}z} &=& \sum_i e^{{\rm i} {\bf Q R}_i} t_{iz}\,.
\end{eqnarray}
Here we have used the fact that a cumulant expectation value does not change when
the operator $e^{\lambda t_{{\bf Q}z}^\dagger}$ is subject to cumulant ordering instead of being applied
directly to the wave function $|\phi_0\rangle$ (see Ref. \onlinecite{Kladko}).
The operator $\Omega e^{\lambda t_{{\bf Q}z}^\dagger}$ can be interpreted as new wave operator
(applied to $|\phi_0\rangle$).
It now contains the additional parameter $\lambda$ which has to be determined
from $
0 = \langle\phi_0| (\news_i^\dagger\newz_i^{})^\cdot\, H \, \Omega\, e^{\lambda t_{{\bf Q}z}^\dagger}|\phi_0\rangle^c$
which is equivalent to eq. (\ref{LAMBDA_EQ}).

\subsection{Ground-state properties}

If we restrict the fluctuation operators in the exponential of $\Omega$ to excitation
pairs as described above (i.e. neglect further operators $S_n$)
and take fully into account the constraint (\ref{CONSTRAINT}),
we find a critical coupling of
$(J_\perp/J_\parallel)_c \sim 3.2$ for the order-disorder transition.
At this level of approximation the magnetization in the limit of decoupled planes
($J_\perp\rightarrow 0$) is around 86 \% of its classical value
being larger than predicted by Monte-Carlo and spin-wave calculations.
The described approximation can be systematically improved by including higher-order fluctuations.
This behavior is also known from calculations within the coupled-cluster method
for the Heisenberg model \cite{Bishop}.
It can be shown that in the limit of an infinite set of operators the cumulant method
yields exact results.
For instance, if we include two additional operators $S_n$ with three and four fluctuations
on adjacent sites into the operator $\Omega$ (\ref{OMEGA_O})
the transition point shifts to $(J_\perp/J_\parallel)_c \sim 2.9$.
However, the analytical and numerical effort to set-up and solve the non-linear
equations equivalent to (\ref{COEFF_EQ_DISO},\ref{COEFF_EQ_O}) increases drastically with including
higher-order fluctuations.

The solution of the non-linear equations (\ref{COEFF_EQ_DISO},\ref{COEFF_EQ_O}) shows that the
absolute values of all coefficients in $\Omega$ are small compared to 1.
In the disordered phase the maximum is taken by $\mu_{10} \sim 0.07$ (the coefficient for
nearest-neighbor triplet pairs) at the point of the phase transition.
In the ordered phase the maximum is at $J_\perp = 0$ with $\mu_{10} \sim 0.14$.
With increasing distance between the triplets the values decay fastly;
higher-order fluctuations acquire a coefficient being one or more orders smaller in
magnitude.
The fact that the coefficients are small is equivalent to the statement that the total triplet density
even at the transition point is small ($\sim 0.1$), see Ref. \onlinecite{Kotov98}.

\begin{figure}
\epsfxsize=8 truecm
\centerline{\epsffile{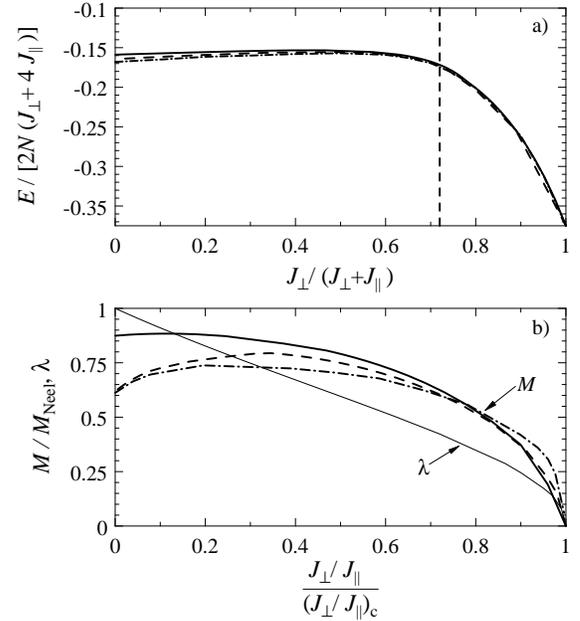}}
\caption{
a) Re-scaled ground-state energy $E$ and b) staggered magnetization $M$
for different approximation levels of the ground-state wave functions
(\protect\ref{OMEGA_DISO}) and (\protect\ref{OMEGA_O}).
Solid: Triplet pairs, no additional operators $S_n$, constraint exact, $(J_\perp/J_\parallel)_c = 3.2$.
Dashed: Triplet pairs, constraint ignored (spin-wave), $(J_\perp/J_\parallel)_c = 4$.
Dash-dot: Series expansion data taken from Ref. \protect\onlinecite{Zheng97} for comparison,
$(J_\perp/J_\parallel)_c = 2.55$.
The vertical dashed line in a) indicates the position of the magnetic quantum phase transition.
In b) the coupling dependence of the basis parameter $\lambda$ is also shown.
}
\label{FIG_GSPROP}
\end{figure}

In Fig. \ref{FIG_GSPROP} we show results for the ground-state energy $E$ and the staggered magnetization $M$
for the approximation using triplet pairs only as described above and for the
spin-wave approach (see below).
The data are compared with recent series expansions from Ref. \onlinecite{Zheng97}.
It can be seen that all approximations well describe the behavior of
the magnetization which first increases with increasing $J_\perp$ and then
drops to zero at the critical coupling.

Another approximation we want to discuss briefly is to
neglect the constraint (\ref{CONSTRAINT}) completely \cite{Chub95,Kotov98} and also the
quartic triplet terms in the Hamitonians (\ref{H_UNDOP_DISO}) and (\ref{H_UNDOP_O}).
The parameter $\lambda$ is simply chosen so that the prefactors of the terms creating
single $\newz$-bosons out of $|\tilde\phi_0\rangle$ vanish.
One can formally set $\news=1$ (condensation of $\news$-bosons);
$|\tilde\phi_0\rangle$ can be considered as "vacuum".
Neglecting the constraint (\ref{CONSTRAINT}) implies that the three types of excitations
$\{\newz_i,t_{ix},t_{iy}\}$ are treated as independent bosons.
After a Fourier transformation one arrives at
\end{multicols}
\widetext
\begin{eqnarray}
H^{(sw)} &=&
  \frac{1}{(1+\lambda^2)^2} \sum_{\bf q}
  ( t_{{\bf q}x}^\dagger t_{{\bf q}x}^{} + t_{{\bf q}y}^\dagger t_{{\bf q}y}^{} )
  \left (
    (1 + \lambda^2) J_\perp  + 8 \lambda^2 J_\parallel + 2 \gamma_{\bf q} (1-\lambda^4) J_\parallel
  \right )
\nonumber \\
&+&
  \frac{1}{(1+\lambda^2)^2} \sum_{\bf q}
   \newz_{\bf q}^\dagger \newz_{\bf q}^{}
  \left (
    (1 - \lambda^4) J_\perp  + 16 \lambda^2 J_\parallel + 2 \gamma_{\bf q} (1-\lambda^2)^2 J_\parallel
  \right )
\nonumber \\
&+&
  J_\parallel \sum_{\bf q} \gamma_{\bf q}
  ( t_{{\bf q}x}^\dagger t_{-{\bf q}x}^{\dagger} + t_{{\bf q}y}^\dagger t_{-{\bf q}y}^{\dagger} + H.c.\;)
\nonumber \\
&+&
  J_\parallel \frac{(1-\lambda^2)^2}{(1+\lambda^2)^2} \sum_{\bf q} \gamma_{\bf q}
   (\newz_{\bf q}^\dagger \newz_{-\bf q}^{\dagger} + H.c.\;)
\label{H_SW}
\end{eqnarray}
\begin{multicols}{2}
\narrowtext
\noindent
with $\gamma_{\bf q} = \frac {1}{2} (\cos q_x + \cos q_y)$ and
\begin{equation}
\lambda = \sqrt{4 J_\parallel - J_\perp \over
                4 J_\parallel + J_\perp}
\end{equation}
in the ordered phase ($4 J_\parallel > J_\perp$) and $\lambda=0$ otherwise.
The Hamiltonian (\ref{H_SW}) can be easily diagonalized by a Bogoliubov transformation.
In this case the ground-state wave functions (\ref{OMEGA_DISO}) and (\ref{OMEGA_O})
(with pairs of triplet excitations) become exact.
The described harmonic approximation can be considered as linear spin-wave theory
for the bilayer problem {\bf with} inclusion
of longitudinal spin fluctuations \cite{Chub95}.
The critical coupling here becomes $(J_\perp/J_\parallel)_c = 4$.
In the limit of decoupled planes the results are equal to that of the linear
spin-wave approach, e.g., the magnetization takes 60.5 \% of its classical
value, see Fig. 2.
Further details on the ground-state calculations will be published elsewhere.


\section{Description of hole motion}

If we remove one electron from rung $i$ a single hole state on this
rung is created. Let us denote by $a^{\dag}_{im,\sigma}$ the creation operator
of a hole with spin $\sigma$ on rung $i$, where the index $m=1,2$ represents
the number of the plane. This operator is defined by the equation
\begin{equation}
\label{a}
a^{\dag}_{im,\sigma}|0\rangle = c^{\dag}_{i{\bar m},\sigma}|0\rangle,
\end{equation}
where ${\bar m}=3-m$. This is a fermionic operator which satisfies the
usual anticommutation relations.
The hole on the rung $i$ interacts with the triplet excitations
on its nearest-neighbor sites. The interaction Hamiltonian can be easily
found by calculating all possible one-hole matrix elements of the initial
Hamiltonian (\ref{H_TJ}) (see Ref. \onlinecite{Eder98}).
For the disordered phase ($\lambda=0$) the result is
\begin{eqnarray}
H_h&=&
-t_\perp\sum_{i\sigma}
  \left(a^{\dag}_{i1,\sigma} \ a_{i2,\sigma}^{} + H.c.\right) \nonumber\\
&+&{t_\parallel \over2}\sum_{\langle ij\rangle m\sigma}
  \left(a^{\dag}_{im,\sigma} \ a_{jm,\sigma}^{} s_j^\dagger s_i^{} +H.c.\right)\nonumber\\
&+&{t_\parallel \over2}\sum_{\langle ij\rangle\sigma}
  \left({\bf t}_{i}^{\dag}{\bf t}_j^{}
  a^{\dag}_{im,\sigma} a_{jm,\sigma}^{} + H.c.\right)\nonumber\\
&+&\sum_{\langle ij\rangle\sigma}
  \left({\bf t}_i^{\dag}\left[
  t_\parallel \ {\bf S}_{i,j} s_j^{}
  +{J_\parallel \over2} \ {\bf S}_{j,j} s_i^{} \right]+H.c. \right)\nonumber\\
&-&t_\parallel \sum_{\langle ij\rangle}
  \left(i \ {\overline {\bf S}}_{j,i}
  \left[{\bf t}_{i}^{\dag}\times {\bf t}_{j}\right] +H.c.\right) \nonumber\\
&-&{J_\parallel \over2}\sum_{\langle ij\rangle}
  \left(i \ {\overline {\bf S}}_{i,i}
  \left[{\bf t}_{j}^{\dag}\times {\bf t}_{j}\right] +H.c.\right).
\label{Hh}
\end{eqnarray}
Following Refs. \onlinecite{Eder98,Sushkov98} we have introduced the notations
${\bf t}=(t_x,t_y,t_z)$ for the triplet vector and
\begin{eqnarray}
\label{s}
&&{\bf S}_{i,j}={1\over2}\sum_{m\alpha\beta}
(-1)^m a^{\dag}_{im,\alpha}{\vec \sigma}_{\alpha\beta} a_{jm,\beta}^{} ,\\
&&{\overline {\bf S}_{i,j}}={1\over2}\sum_{m\alpha\beta}
a^{\dagger}_{im,\alpha}{\vec\sigma}_{\alpha\beta} a_{jm,\beta}^{}.
\nonumber
\end{eqnarray}
for generalized hopping operators
where ${\vec \sigma}$ is the vector of Pauli matrices.
In addition we have to impose the constraint that a hole state and one of
the doubly occupied states cannot coexist on the same rung:
\begin{equation}
\label{c2}
s_i^{\dagger} s_i^{} + \sum_\alpha t_{i\alpha}^{\dagger}t_{i\alpha}^{}
                     + \sum_{m\sigma} a_{im,\sigma}^{\dagger}a_{im,\sigma}^{} = 1.
\end{equation}
The several terms in the Hamiltonian (\ref{Hh}) have been discussed by Eder \cite{Eder98}.
They contain hopping without changing the spin background
(direct hopping, second and third term) as well as spin-fluctuation assisted
hopping (fourth and fifth term) and exchange processes where the
hole remains on its rung (last two terms).
Together with the Hamiltonian for the doubly occupied rungs (\ref{H_UNDOP_DISO}) the
Hamiltonian (\ref{Hh}) contains the complete one-hole dynamics
for the disordered phase.
For the case of more than one hole additional hole interaction processes would
have to be considered. However, in the present work we restrict ourselves
to the discussion of the one-hole problem.

If we work in the magnetically ordered phase it is convenient to use
the generalized basis $\{\news,\newz,t_x,t_y\}$ from Sec. II to represent
the states of the doubly occupied rungs.
From the Hamiltonian (\ref{Hh}) and the definition (\ref{BASIS_DEF})
of the basis operators one can construct a Hamiltonian for an ordered
background state.
It again contains processes proportional to $t_\parallel$ where the hole hops to
a neighboring rung and exchange terms propertional to $J_\parallel$.
To be short we only state some hopping matrix elements of the resulting
Hamiltonian.
For that we symbolically introduce two-rung states
$|XY\rangle=X_i^\dagger Y_j^\dagger|0\rangle$
where $X$ and $Y$ are the states on two adjacent rungs
with $i\in A$ and $j\in B$.
\begin{eqnarray}
\langle \news a_{0\uparrow} | H_h | a_{0\uparrow} \news \rangle \;&=&\;
  {t_\parallel \over 2} {{1-\lambda^2} \over {1+\lambda^2}} \,,
\nonumber\\
\langle \newz a_{0\uparrow} | H_h | a_{0\uparrow} \newz \rangle \;&=&\;
  {t_\parallel \over 2} {{1-\lambda^2} \over {1+\lambda^2}} \,,
\nonumber\\
\langle \newz a_{0\uparrow} | H_h | a_{0\uparrow} \news \rangle \;&=&\;
  -{t_\parallel \over 2} {{(1+\lambda)^2} \over {1+\lambda^2}} \,,
\nonumber\\
\langle t_x a_{0\downarrow} | H_h | a_{0\uparrow} \news \rangle \;&=&\;
  -{t_\parallel \over 2} {{1+\lambda} \over \sqrt{1+\lambda^2}} \,.
\label{HOLE_MATEL}
\end{eqnarray}
The first two terms represent processes of hole motion without desturbing the
spin background (so-called direct hopping) whereas the last two terms are examples for
hole hopping with creation of a "spin defect".
It can be seen that direct hopping is impossible in the limit of $|\lambda|=1$, i.e.,
in a N\'{e}el-ordered background.


\section{Single hole dynamics}

To investigate the hole motion we consider a one-particle Green's function
describing the creation of a single hole with momentum $\bf k$
at zero temperature:
\begin{equation}
G({\bf k},\omega) \;=\;
  \langle\psi_0| {\hat c}_{{\bf k}\sigma}^\dagger {1 \over {z-L}}
               {\hat c}_{{\bf k}\sigma}
  |\psi_0\rangle
\label{HOLE_GF}
\end{equation}
where $z$ is the complex frequency variable, $z=\omega+i\eta$,
$\eta\rightarrow 0$.
The quantity $L$ denotes the Liouville operator defined by $LA = [H,A]_-$ for arbitrary
operators $A$;
${\bf k}=(k_x,k_y,k_z)$ is the hole momentum with $k_z = 0$ or $\pi$ for
the bonding or antibonding band, respectively.
$|\psi_0\rangle$ is the full ground state of the undoped system
as described in Sec. II.
The effect of the hole creation operator ${\hat c}_{im,\sigma}$ applied to $|\psi_0\rangle$
can be related to the fermion operators $a_{im,\sigma}^\dagger$ introduced in the last section.
One obtains for example for $i \in A$ sublattice
\begin{eqnarray}
{\hat c}_{i1,\uparrow} \, \news_i^\dagger |0\rangle
&=&
{1+\lambda \over \sqrt{2(1+\lambda^2)}} \, a_{i1,\downarrow}^\dagger |0\rangle  \,,
\nonumber \\
{\hat c}_{i1,\uparrow} \, t_{ix}^\dagger |0\rangle
&=&
- {1 \over \sqrt{2}} \, a_{i1,\uparrow}^\dagger |0\rangle
\,,
\end{eqnarray}
and analogous relations for other basis states and for $i \in B$.

The hole motion processes will be described in the concept of path operators
~\cite{Strings,Trugman88,Vojta98}
which create strings of spin fluctuations attached to the hole.
For the application of projection technique we define a set of path operators
$\{A_I\}$ which couple to a hole and create local spin defects with respect to
the undoped ground state $|\psi_0\rangle$.
The first operator $A_0$ is the unity operator, the second one $A_1$ moves the hole
by one lattice spacing creating one spin excitation and so on.
We are interested in calculating dynamical correlation functions for the
operators $\{A_I c_{{\bf k}\sigma}\}$:
\begin{equation}
G_{I\sigma,J\sigma'}(z)
       \,=\, \langle\psi_0|\,(A_I c_{{\bf k}\sigma})^\dagger {1 \over {z-L}}
                               (A_J c_{{\bf k}\sigma'})^{\phantom\dagger}\,
               |\psi_0\rangle \,.
\label{ARBKORR}
\end{equation}
The Green's function (\ref{HOLE_GF}) for the hole is then
given by $G_{0\sigma,0\sigma}(z)$.
Using cumulants the correlation functions $G$ can be rewritten as \cite{BeckBre90}:
\begin{equation}
G_{I\sigma,J\sigma'} (z) \,=\,
  \langle\phi_0|\,\Omega^\dagger\,
  (A_I c_{{\bf k}\sigma})^\dagger \left( {1 \over {z-L}}
  A_J c_{{\bf k}\sigma'}^{\phantom\dagger} \right)^{\cdot} \, \Omega\,
  |\phi_0\rangle^c
\,.
\label{KUMKORR}
\end{equation}
The brackets $\langle\phi_0|\,...\,|\phi_0\rangle^c$ denote cumulant expectation
values with $|\phi_0\rangle$.
%
The "unperturbed" ground state at half-filling, i.e., one of the product
states (\ref{SINGLET_PROD}) or (\ref{CONDENS_PROD}) for the
disordered or ordered system, respectively, will be denoted in the following by
$|\phi_0\rangle$ for both phases.
The operator $\Omega$ has been described above, it transforms $|\phi_0\rangle$
being the ground state of $H_0$ into the full ground state $|\psi_0\rangle$ of
$H = H_0+H_1$ at half filling.
In the following, $\Omega$ is approximated by an exponential
ansatz as described in Sec. II.
Since the expression (\ref{KUMKORR}) contains both $\Omega$ and $\Omega^\dagger$,
the expectation values can be calculated only in the lowest non-trivial order of the
fluctuations in $\Omega$, i.e., the expansion will be restricted to
linear fluctuation terms.
In order to correctly describe the short-range spin fluctuations in the system,
operators with up to 4 triplet excitations with a maximum
distance of 4 lattice spacings have been employed.
For the evaluation of their coefficients we have linearized the
set of equations (\ref{COEFF_EQ_DISO},\ref{COEFF_EQ_O}).
This is possible since the absolute values of the coefficients are small compared to 1
as discussed in Sec. II C.
Therefore quadratic and higher terms in the fluctuations operators can be
neglected.
By comparing the coefficients of the short-range fluctuations obtained in this way
with the ones from the full (non-linear) ground-state calculation of Sec. II C
we have numerically verified that the error in the matrix elements (\ref{QAFM_MATEL})
introduced by this linearization is smaller than a few percent.
Since the linearized equations for the coefficients do not yield a phase transition point
we fix it at the value $(J_\perp/J_\parallel)_c = 2.55$ known from series
expansions \cite{Zheng97} and Monte Carlo calculations \cite{SaSca}.
(In the limit of an infinite set of operators $\{S_n\}$ within the cumulant method the solution
of the non-linear equations has to yield the "exact" transition point which we assume to be
at $(J_\perp/J_\parallel)_c = 2.55$.)
In the ordered phase we set
\begin{equation}
\lambda = \sqrt{(J_\perp/J_\parallel)_c - J_\perp/J_\parallel \over
               (J_\perp/J_\parallel)_c + J_\perp/J_\parallel}
\end{equation}
which is suggested by the results obtained in Sec. II.

Using Mori-Zwanzig projection technique \cite{Projtech} one can derive
a set of equations of motion for the dynamical correlation
functions $G_{I\sigma,J\sigma'}(z)$.
Neglecting the self-energy terms it reads:
\begin{eqnarray}
\sum_{I\sigma}  \Omega_{K\tilde\sigma,I\sigma} (z)
  \,G_{I\sigma,J\sigma'}(z)\,\,=\,\,\chi_{K\tilde\sigma,J\sigma'} \, ,
  \nonumber \\
\Omega_{K\tilde\sigma,J\sigma'}(z) =
  z \delta_{KJ}\delta_{\tilde\sigma\sigma'}
  - \sum_{L\sigma''}\omega_{K\tilde\sigma,L\sigma''}^{\phantom{-1}}\,\chi_{L\sigma'',J\sigma'}^{-1}
\,.
\label{PROJ_GLSYS}
\end{eqnarray}
$\chi_{I\sigma,J\sigma'}$ and $\omega_{I\sigma,J\sigma'}$
are the static correlation functions and frequency terms,
respectively.
They are given by the following cumulant expressions:
\begin{eqnarray}
\chi_{I\sigma,J\sigma'} &=&
  \langle\phi_0| \Omega^\dagger
  (A_I c_{{\bf k}\sigma})^{\cdot \dagger}
  (A_J c_{{\bf k}\sigma'})^\cdot \Omega \,
  |\phi_0\rangle^c \, , \nonumber\\
\omega_{I\sigma,J\sigma'} &=&
  \langle\phi_0| \Omega^\dagger
  (A_I c_{{\bf k}\sigma})^{\cdot \dagger}
  \left(L (A_J c_{{\bf k}\sigma'})^{\phantom\dagger} \right)^\cdot \Omega \,
  |\phi_0\rangle^c
\,.
\label{QAFM_MATEL}
\end{eqnarray}
These terms describe all dynamic processes within the subspace of the
Liouville space spanned by the operators $\{A_I c_{{\bf k}\sigma}\}$.
The use of cumulants ensures size-consistency, i.e., only spin fluctuations connected
with the hole enter the final expressions for the one-hole correlation
function.

In the present calculations we have employed up to 1600 projection variables with
a maximum path length of 3.
The neglect of the self-energy terms leads to a discrete set of poles for
the Green's functions, so the present approach cannot account for
linewidths.
In all figures we have introduced an artificial linewidth of $0.2 t_\parallel$
to plot the spectra.
For details of the calculational procedure see e.g. Ref. \onlinecite{Vojta98}.

\subsection{One-hole spectrum}

Now we turn to the discussion of the final results.
First we consider the case of vanishing interplane hopping $t_\perp=0$, i.e.,
the hole motion is restricted to one plane.
(In this case $k_z$ can be dropped.)

\begin{figure}
\epsfxsize=8.3 truecm
\centerline{\epsffile{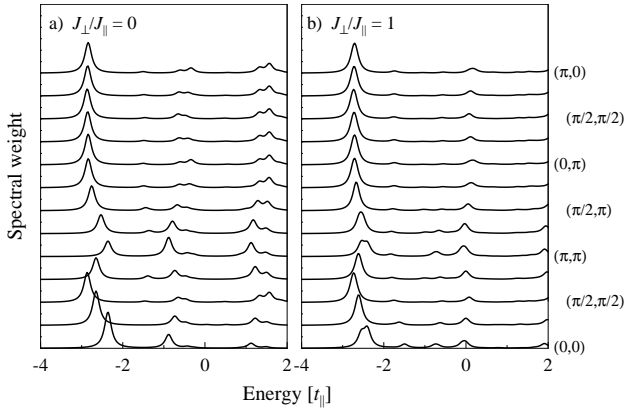}}
\caption{
One-hole spectral function for $t_\perp=0$ (no interplane hopping),
a) $J_\perp/J_\parallel=0$, b) $J_\perp/J_\parallel=1$, and
different momenta $(k_x,k_y)$.
The other parameters are $J_\parallel/t_\parallel=0.4$; the energies
are measured in units of $t_\parallel$ relative to the energy of
a localized hole.
}
\label{FIG_SPEC0_1}
\end{figure}

\begin{figure}
\epsfxsize=8.3 truecm
\centerline{\epsffile{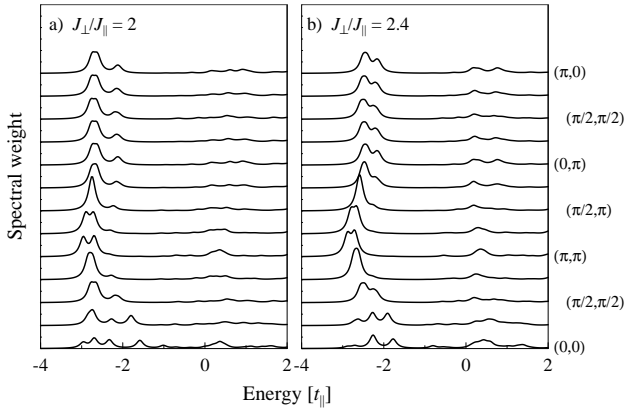}}
\caption{
Same as Fig. \protect\ref{FIG_SPEC0_1}, but for
a) $J_\perp/J_\parallel=2$, b) $J_\perp/J_\parallel=2.4$.
}
\label{FIG_SPEC0_2}
\end{figure}

\begin{figure}
\epsfxsize=8.3 truecm
\centerline{\epsffile{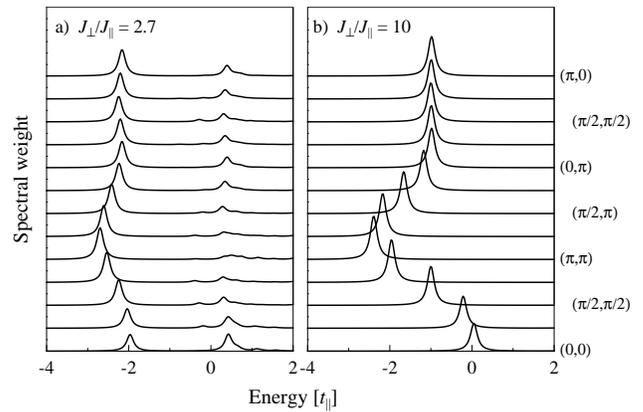}}
\caption{
Same as Fig. \protect\ref{FIG_SPEC0_1}, but for
a) $J_\perp/J_\parallel=2.7$, b) $J_\perp/J_\parallel=10$.
}
\label{FIG_SPEC0_3}
\end{figure}

The one-hole spectral function ${\rm Im}\,G({\bf k},\omega)$
for $t_\perp=0$, $J_\parallel/t_\parallel=0.4$ and various values of $J_\perp/J_\parallel$ is shown in
Figs. \ref{FIG_SPEC0_1} -  \ref{FIG_SPEC0_3}.
For zero and small $J_\perp$, i.e., in the limit of decoupled planes, we observe
a pronounced quasiparticle (QP) peak and a weak background at higher energies.
The QP peak follows a dispersion with minima at $(\pm\pi/2,\pm\pi/2)$.
These features are well-known from the one-hole problem in the single-layer
antiferromagnet.
With increasing $J_\perp$ the dispersion is "washed-out" because the antiferromagnetic
correlations in the background state are weakened.
At $J_\perp/J_\parallel \sim 1.5$ a cross-over to a dispersion with minima at
$(\pi,\pi)$ and $(0,0)$ occurs.
Note that this cross-over point still lies {\bf inside} the antiferromagnet phase.
Near the phase transition the spectral weight of the lowest pole at $(0,0)$
becomes small.
Most of the spectral weight can be found in a band with minimum at $(\pi,\pi)$ and
maximum at $(0,0)$.
The weak band visible at the bottom of the spectrum around momentum $(0,0)$ can be
considered as "shadow band" originating from
the antiferromagnetic background, i.e., it is obtained by shifting the band with large
weight by ${\bf Q} = (\pi,\pi)$.

Further increasing $J_\perp$ drives the system into the gapped paramagnetic (PM) phase.
At the phase transition it can be seen that shadow bands being present in the AF phase
disappear in the PM phase. At large $J_\perp$ the hole behaves like a free
fermion with a dispersion proportional to $t_\parallel (\cos k_x + \cos k_y)$.

\begin{figure}
\epsfxsize=8.3 truecm
\centerline{\epsffile{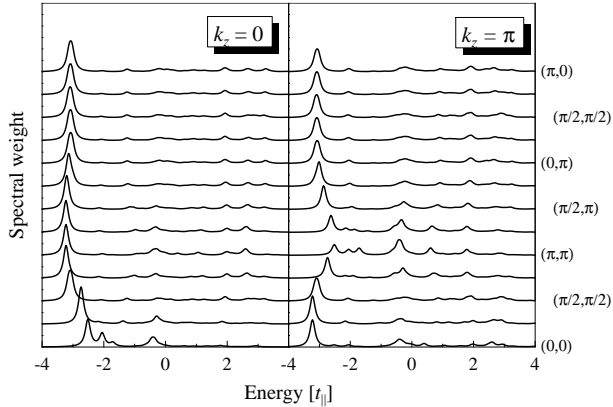}}
\caption{
One-hole spectral function for $t_\perp/t_\parallel = 1$,
$J_\perp/J_\parallel=1$, $J_\parallel/t_\parallel=0.4$, and
different momenta.
The left and right panel show the bonding ($k_z=0$) and antibonding ($k_z=\pi$)
bands, respectively.
}
\label{FIG_SPEC100_U25}
\end{figure}

\begin{figure}
\epsfxsize=8.3 truecm
\centerline{\epsffile{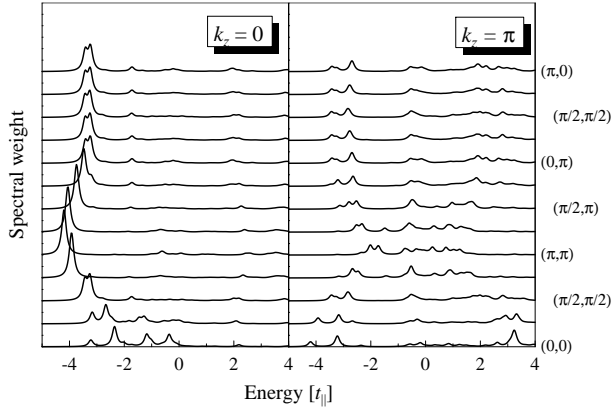}}
\caption{
Same as Fig. \protect\ref{FIG_SPEC100_U25},
but for $t_\perp/t_\parallel = 1.5$,
$J_\perp/J_\parallel=2.25$, and $J_\parallel/t_\parallel=0.4$.
}
\label{FIG_SPEC225_U25}
\end{figure}

\begin{figure}
\epsfxsize=8.3 truecm
\centerline{\epsffile{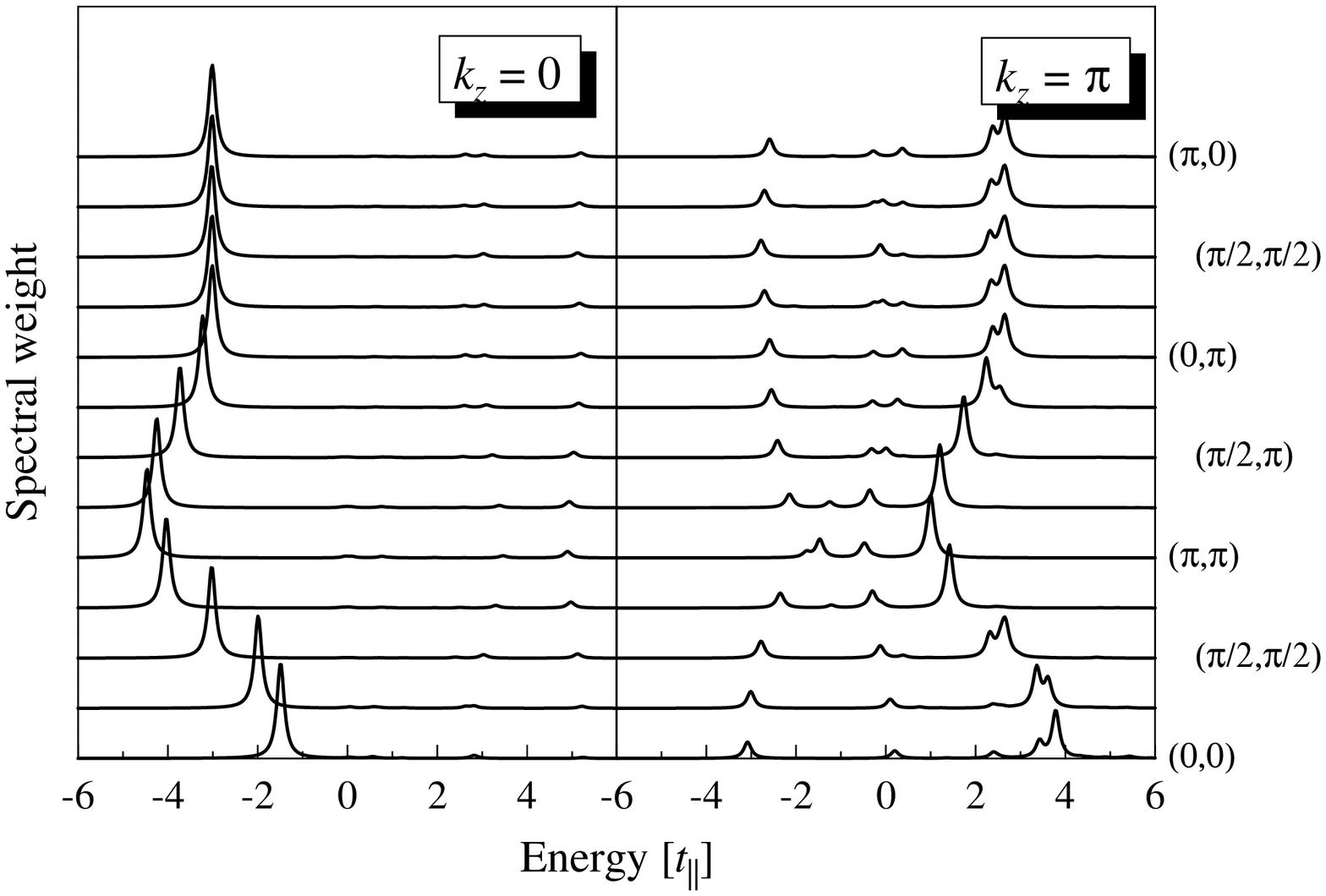}}
\caption{
Same as Fig. \protect\ref{FIG_SPEC100_U25},
but for $t_\perp/t_\parallel = 2$,
$J_\perp/J_\parallel=4$, and $J_\parallel/t_\parallel=0.4$.
}
\label{FIG_SPEC400_U25}
\end{figure}

Now we consider the effect of interplane hopping.
We fix the ratio of the parameters to
\begin{equation}
{t_\perp^2 \over J_\perp} \:=\: {t_\parallel^2 \over J_\parallel} \:=\: {U \over 4}
\end{equation}
which follows from the derivation of the $t-J$ model from a Hubbard model for the bilayer system
with on-site repulsion $U$.
Spectra for different parameter sets with $U/t_\parallel = 10$ are shown in
Figs. \ref{FIG_SPEC100_U25} - \ref{FIG_SPEC400_U25}.
An interplane coupling of $t_\perp/t_\parallel=1$ has already changed the character
of the bands compared to $t_\perp=0$ (cf. Fig. \ref{FIG_SPEC0_1}a).
Further increasing $t_\perp/t_\parallel$ to 1.5 (Fig. \ref{FIG_SPEC225_U25}) forms a pronounced band
with a free-fermion form at the bottom of the spectrum in the bonding channel ($k_z = 0$).
In contrast, the antibonding spectrum ($k_z=\pi$)
becomes incoherent since spectral weight is transferred to higher energies.
Since we are still in the antiferromagnetic phase the lowest peaks in the antibonding
channel appear at the same energies as in the bonding channel shifted by momentum
$\bf Q = (\pi,\pi,\pi)$, i.e., these can again be considered as shadow bands.
At $t_\perp/t_\parallel > \sqrt{2.55}$ the AF order is destroyed.
As shown in Fig. \ref{FIG_SPEC400_U25} for $t_\perp/t_\parallel = 2$ the shadow bands are
disappeared, and in the antibonding channel a pronounced free-fermion band at high
energies is formed.
Higher values of $t_\perp$ will completely suppress the incoherent weight left in the antibonding
channel.

Varying the ratio $J_\parallel/t_\parallel$ (which is 0.4 for all figures presented here)
does not change the picture qualitatively. Larger values of $J_\parallel$ suppress the
incoherent background in all spectra. For very small values of $J_\parallel$ the spectral
functions become incoherent since the hole is dressed with an increasing number of spin
fluctuations, i.e., the radius of the spin polaron increases.
Furthermore, the bandwidth of the quasiparticle dispersion at small $J_\perp$ is mainly controlled by
$J_\parallel$ (instead of $t_\parallel$).
This means that the bandwidth in units of $t_\parallel$ decreases with
decreasing $J_\parallel/t_\parallel$, see also next subsection.
The location of the cross-over between the two dispersion forms depends weakly on
$J_\parallel/t_\parallel$ which will be discussed below.

\subsection{Quasiparticle bands}

\begin{figure}
\epsfxsize=6.5 truecm
\centerline{\epsffile{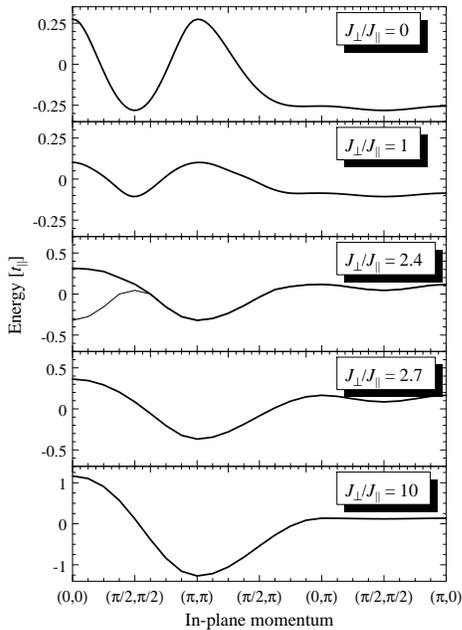}}
\caption{
Quasiparticle dispersion for $t_\perp=0$ (no interplane hopping),
$J_\parallel/t_\parallel = 0.4$,
and different values of $J_\perp/J_\parallel$.
The energy zero level has been set at the center of mass of the band.
For $J_\perp/J_\parallel = 2.4$ the heavy line shows the dispersion of
the peak carrying the main spectral weight whereas the thin line
corresponds to the lowest pole in the spectrum (shadow band, see text).
}
\label{FIG_DISP1}
\end{figure}

Next we are going to examine the properties of the low-lying bands.
Again we first discuss the case of vanishing interplane hopping,
$t_\perp=0$.
For $J_\perp \ll J_\parallel$ one finds a narrow QP band shown in Fig. \ref{FIG_DISP1}
(top panel).
It has minima at $(\pm\pi/2,\pm\pi/2)$ and a bandwidth of approximately
$1.5 J_\parallel$.
It corresponds to the coherent motion of a dressed hole, i.e., a spin-bag
quasiparticle.
(Note that a larger set of path operators would give a larger bandwidth of around
$2 J_\parallel$ which is known from the calculations for a single-layer antiferromagnet
\cite{Vojta98}. However, paths of length 4 and more are hard to access for the bilayer
system due to the increasing numerical effort.)

The main contribution to the hole motion in the AF case can be understood as
follows: the hopping hole locally destroys the antiferromagnetic spin order
leaving behind a string of spin defects. Quantum spin fluctuations
can repair pairs of frustrated spins which leads to a coherent motion
in one of the antiferromagnetic sublattices
(spin-fluctuation-assisted hopping).
For $t_\parallel/J_\parallel > 1$ the bandwidth of this coherent hole motion is of order
$J_\parallel$ because the spin-flip part of the intraplane Heisenberg interaction $J_\parallel$
is necessary to remove the spin defects caused by hopping.

With increasing $J_\perp$ the bandwidth of the dispersion first decreases;
at a value of $J_\perp/J_\parallel \sim 1.5$ the minima move to $(\pi,\pi)$ and $(0,0)$.
Slightly larger values of $J_\perp$ lead to an increase of the bandwidth,
and spectral weight is transferred to a band which has nearly
tight-binding form.
So the lowest pole at $J_\perp/J_\parallel = 2.4$
has very small weight around momentum $(0,0)$.
This shadow band is shown as dashed line in Fig. \ref{FIG_DISP1}.
In the paramagnetic phase for larger $J_\perp$ the shadow band disappears.
The QP peak has a dispersion being nearly the one of a free fermion but with
a reduced bandwidth.
This behavior follows from the existence of the direct hopping term
with prefactor $t_\parallel/2$ in $H_h$ (\ref{Hh}).
It leads to a disperison proportional to $t_\parallel$; the saturation bandwidth
equals half of the bandwidth of the uncorrelated system ($8\,t_\parallel$).

\begin{figure}
\epsfxsize=6.5 truecm
\centerline{\epsffile{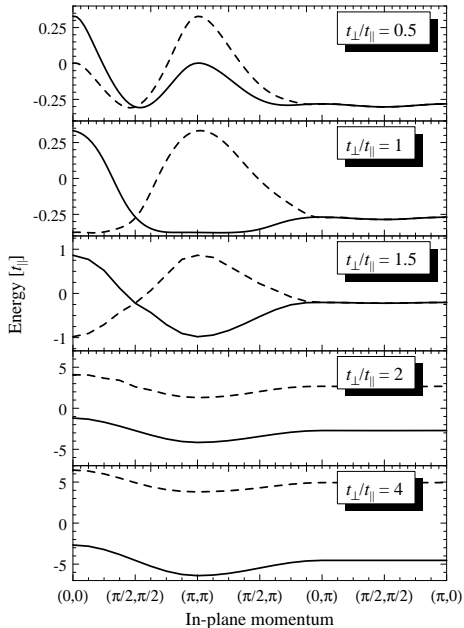}}
\caption{
Quasiparticle dispersion for
${t_\perp^2 \over J_\perp} \:=\: {t_\parallel^2 \over J_\parallel} \:=\: {U \over 4}$
with $U/t_\parallel=10$ and different values of $t_\perp/t_\parallel$.
The in-plane momentum $(k_x,k_y)$ varies along the horizontal axis.
Solid/dashed lines correspond to bonding/antibonding modes ($k_z=0$ or $\pi$).
}
\label{FIG_DISP2}
\end{figure}

The results for non-zero interplane hopping $t_\perp$ are shown in Fig. \ref{FIG_DISP2}.
Again, for small $t_\perp$ (small $J_\perp$) the dispersion character is
"antiferromagnetic" whereas for large $t_\perp$ (large $J_\perp$)
the bands have a simple tight-binding form.
For $J_\perp \gg J_\parallel$ and $t_\perp \gg t_\parallel$
their dispersion is given by
\begin{equation}
\epsilon_{\bf k} = - t_\perp \cos k_z  + t_\parallel (\cos k_x + \cos k_y)
\end{equation}
The bandwidth for the in-plane motion is reduced by a factor of 2 compared to
a free fermion since the relevant hopping matrix element is
$\langle s a_{0\uparrow} | H_h | a_{0\uparrow} s \rangle = t_\parallel/2$.

\begin{figure}
\epsfxsize=8 truecm
\centerline{\epsffile{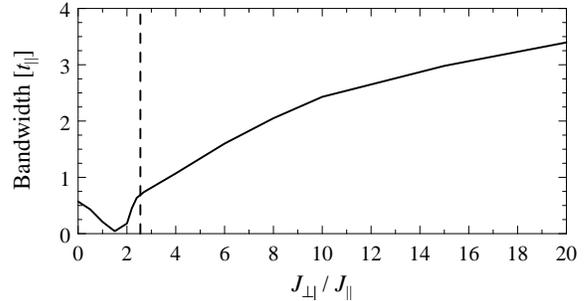}}
\caption{
Bandwidth of the quasiparticle dispersion (shown in Fig. \ref{FIG_DISP1})
vs. $J_\perp/J_\parallel$
for $t_\perp=0$ (no interplane hopping) and $J_\parallel/t_\parallel = 0.4$.
The vertical dashed line again indicates the position of the magnetic phase
transition.
}
\label{FIG_BANDW1}
\end{figure}

In Fig. \ref{FIG_BANDW1} we finally show the bandwidth of the QP dispersion for the case of
vanishing interplane hopping and $J_\parallel / t_\parallel = 0.4$, i.e., for the dispersions shown
in Fig. \ref{FIG_DISP1}. It can clearly be seen that the cross-over from an "antiferromagnetic" to a
simple tight-binding dispersion occurs at $J_\perp/J_\parallel \sim 1.5$ and is connected with a
pronounced minimum in the QP bandwidth. For $J_\perp/J_\parallel > 1.5$ the bandwidth increases and
saturates for large $J_\perp$ at $4 t_\parallel$ as discussed above.
Results for other values of
$J_\parallel/t_\parallel$ are qualitatively similar. The minimum in the dispersion
(i.e. the cross-over point) shifts with
$J_\parallel/t_\parallel$: for $J_\parallel/t_\parallel = 0.1$ it is located at
$J_\perp/J_\parallel \sim 1.1$, for $J_\parallel/t_\parallel = 10$ it is found at
$J_\perp/J_\parallel \sim 2.0$.
This in turn means that the cross-over in the shape of the single-hole dispersion can
be driven by varying $J_\parallel/t_\parallel$ at {\bf fixed} $J_\perp/J_\parallel$.

\subsection{Relation to the short-range spin correlations}

In order to understand the behavior of the QP dispersion and its bandwidth
we illustrate in more detail the connection between the in-plane hole motion and the
short-range in-plane spin correlations.
The basic ingredience for the observed behavior of the QP dispersion is the
competition between direct hopping which dominates in the disordered
phase and spin-fluctuation-assisted hopping known from
the single-layer AF.
The relevant matrix element for direct hopping between states without spin deviations is
$t_1 := \langle\psi_0| {\hat c}_{j\sigma}^\dagger H_t {\hat c}_{i\sigma} |\psi_0\rangle$
where $i,j$ are nearest-neighbor sites and $|\psi_0\rangle$ denotes the undoped
background state.
This term can be expressed \cite{Vojta98,HopCorr} by the static
in-plane nearest-neighbor spin correlation
$S_{\bf R} = \langle\psi_0|{\bf S}_0 \cdot {\bf S}_{\bf R} |\psi_0\rangle$ with ${\bf R}=(1,0)$
as
\begin{equation}
t_1 =
2 t_\parallel ( S_{10} + {1 \over 4}) \,.
\end{equation}
Spin-fluctuation-assisted hopping is more complicated since it involves two hopping steps
and one spin-fluctuation process.
The matrix element for the spin-fluctuation process (being the
most important for not too small values of $t_\parallel$) can be written as
\mbox{$t_2 := \langle\psi_0|{\hat c}_{j\sigma}^\dagger H_J A_2 {\hat c}_{i\sigma}|\psi_0\rangle$} where
$A_2$ moves the hole by two hopping steps (creating a path of spin defects) and $i,j$ are now
next-nearest neighbors.
Transforming the expectation value into spin correlation functions one obtains \cite{Vojta98,HopCorr}
\begin{equation}
t_2 =
{J_\parallel} ( -{1 \over 2} S_{10} + {S_{20} + 2 S_{11} \over 4} + {3 \over 16})
\end{equation}
where the average over the possible paths of length 2 has already been performed.

\begin{figure}
\epsfxsize=7.5 truecm
\centerline{\epsffile{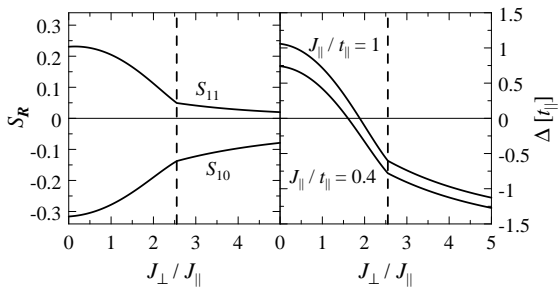}}
\caption{
Left: $J_\perp/J_\parallel$-dependence of the
equal-time short-range spin correlation functions $S_{\bf R}$ in the spin background
state calculated with the linearized exponential ansatz used for the
description of the hole motion.
Right:
$\Delta$ vs. $J_\perp/J_\parallel$ from eq. (\protect\ref{DELTA_APPROX})
for $J_\parallel/t_\parallel = 0.4$ and 1.0.
}
\label{FIG_SPINCORR1}
\end{figure}

The in-plane dispersion shape can be described by the energy difference
$\Delta = E(\pi,\pi) - E(\pi/2,\pi/2)$.
Values $\Delta > 0$ correspond to an "antiferromagnetic" dispersion whereas
$\Delta < 0$ occurs for a nearest-neighbor tight-binding dispersion.
Neglecting longer paths and more complicated contributions to the hole motion
the QP dispersion is given by the sum of a nearest-neighbor and a next-nearest neighbor dispersion
originating from the two processes described above.
Then $\Delta$ can be roughly estimated from the matrix elements $t_1$ and $t_2$:
\begin{equation}
\Delta \sim - 4 t_1 + t_2 \,.
\label{DELTA_APPROX}
\end{equation}
The prefactor of $t_1$ arises from the number of nearest neighbors;
for the $t_2$-prefactor the influence of two hopping steps has to be kept
in mind, a fit to numerical results for the single-layer problem yields a value
of order 1.
From this we see that the cross-over phenomenon can be understood in terms of the interplay
between short-range spin correlations and the ratio of $J_\parallel/t_\parallel$.

Fig. \ref{FIG_SPINCORR1} shows the values of $S_{\bf R}$ for nearest and next-nearest
neighbor sites obtained from the present calculation.
With increasing $J_\perp/J_\parallel$ the absolute values of the in-plane spin correlation
functions decrease from their single-layer values and are weakened within the
antiferromagnetic phase.
At the transition point the magnitude has dropped by a factor around $2-3$ compared to
$J_\perp=0$ which again coincides with the fact that the density of spin (triplet) excitations
is small in the disordered phase even at the transition point \cite{Kotov98}.
We have also plotted the quantity $\Delta$ estimated from eq. (\ref{DELTA_APPROX})
for two different values of $J_\parallel/t_\parallel$.
One notices a semi-quantitative agreement of the zero in $\Delta$ and the dispersion minimum
described in Sec. IV B.


\section{Conclusions}

In this paper we have presented for the first time a systematic analytical study of the
one-hole dynamics on both sides of a magnetic ordering transition in a low-dimensional
antiferromagnet.
The system under consideration was a bilayer antiferromagnet described by a
$t$-$J$ Hamiltonian in the limit of zero doping.
The magnetic background state has been modelled with modified bond
operators.
In the disordered phase these operators describe the singlet ground state
and the triplet excitations.
In the AF ordered phase they account for the condensation of one type of triplet
bosons and describe transverse as well as longitudinal fluctuations.
Using the spin-polaron concept which describes the one-hole states in
terms of local spin deviations we have calculated the one-hole spectral
function for the whole range of magnetic couplings.

For the disordered background (large $J_\perp/J_\parallel$, gapped spin excitations)
spin fluctuations around the hole are suppressed. The hole motion is dominated by direct
hopping processes, i.e., hopping without disturbing the spin
background.
In the ordered phase for very small interplane coupling we recover the
results known from the single-layer hole motion: The spectrum consists
of a coherent QP peak at the bottom and an incoherent background.
The QP can be associated with a mobile hole dressed by spin fluctuations.
The bandwidth of its coherent motion is controlled by
$J_\parallel$.

The cross-over between these two scenarios occurs {\bf inside} the ordered phase
where the antiferromagnetic short-range correlations become weakened.
The cross-over is located between $1 < J_\perp/J_\parallel < 2$ depending on $J_\parallel /t_\parallel$
(for $J_\parallel /t_\parallel = 0.4$ it is found at $J_\perp/J_\parallel \sim 1.5$).
Note that the cross-over can also be driven by variation of the hopping strength
$t_\parallel$ at a fixed value of $J_\perp/J_\parallel$ well in the antiferromagnetic
phase (e.g. 1.3).
This behavior follows from the competition between direct nearest-neighbor hopping and
spin-fluctuation-assisted next-nearest-neighbor hopping.
In contrast, when crossing the magnetic phase boundary at $J_\perp / J_\parallel \sim 2.5$
there are no drastic changes in the
spectrum (and therefore in the ARPES response of such a system).
The only differences between the spectra in both phases near the phase transition are weak shadow
bands in some regions of the Brillouin zone in the AF phase.

Note that our approximation which describes the hole motion processes in terms
of short-range spin fluctuations is questionable in a region in the close vicinity
of the phase transition due to the existence of long-ranged critical
fluctuations.
However, it has been argued recently\cite{Sushkov98} that the influence of
the critical modes suppresses the quasiparticle weight only in the limit
of vanishing hopping $t/J \rightarrow 0$.
In contrast, for finite hopping the number of spin fluctuations near the hole remains
finite even at the transition point.
So we expect the picture presented in this paper to be valid at least in the regions away
from the transition, i.e., $|J_\perp / J_\parallel - 2.55| > 0.1$,
which is supported by the fact that our spectral functions on both sides of the phase
transition (e.g. at $J_\perp / J_\parallel = 2.4$ and 2.7) do not show major
differences.

So the main result of the present paper can be summarized as follows:
A magnetic phase transition
has only weak influence on the ARPES spectrum of a doped antiferromagnet.
The properties of the spectrum are, however, dominated by the short-range
environment of the hole.
The statement is expected to hold also for the doping-induced phase
transition in the single-layer AF.
Here the antiferromagnetic correlations are strong even in the paramagnetic
phase, i.e., from the above considerations one expects to find
an "antiferromagnetic" hole dispersion on both sides of the
transition.
This is exactly what was observed in recent work \cite{Dagotto94,Brenig95}:
analytical investigations of one hole in the AF phase as well as numerical
studies of finite systems (which have a singlet ground state without long-range
order) both show a coherent hole motion with a dispersion of width $2 J$ and minima at
$(\pm\pi/2,\pm\pi/2)$.

It should be pointed out that the above discussion applies to intermediate and high
energy scales (order $t$, $J$) only.
Of course there exist low-energy properties of the spectrum which are expected to
be influenced by quantum criticality, e.g., the linewidth in a finite-temperature
photoemission experiment should show scaling behavior in the quantum-critical
region associated with the transition \cite{valla}.
These features as well as the hole dynamics in the bilayer system at low but finite doping
are beyond the scope of the present study and will be subject of future research.

\acknowledgments
{
The authors thank E. Dagotto and T. Sommer for useful conversations.
M.V. acknowledges support by the DFG (VO 794/1-1).
}



\end{multicols}

\end{document}